%% file: conference.tex
\documentclass[10pt, conference]{article}
\usepackage[noadjust]{cite}
\usepackage{amsmath, amssymb, amsfonts, url, spconf}
\usepackage[ruled,vlined]{algorithm2e}
\usepackage{graphicx, tabularx, booktabs, subcaption}
\usepackage{textcomp}
\usepackage{xcolor}
\usepackage[acronym]{glossaries}

\DeclareMathOperator*{\argmax}{arg\max}
\def\BibTeX{{\rm B\kern-.05em{\sc i\kern-.025em b}\kern-.08em
    T\kern-.1667em\lower.7ex\hbox{E}\kern-.125emX}}

\begin{document}

\title{Low-Complexity Maximum Likelihood Detection for Type-Based Over-the-Air Computation
}

% \author{\IEEEauthorblockN{Marc Martinez-Gost\IEEEauthorrefmark{1}\IEEEauthorrefmark{2}, Ana Pérez-Neira\IEEEauthorrefmark{1}\IEEEauthorrefmark{2}\IEEEauthorrefmark{3}, 
% Miguel Ángel Lagunas\IEEEauthorrefmark{2}}
% \IEEEauthorblockA{
% \IEEEauthorrefmark{1}Centre Tecnològic de Telecomunicacions de Catalunya, Spain\\
% \IEEEauthorrefmark{2}Dept. of Signal Theory and Communications, Universitat Politècnica de Catalunya, Spain\\
% \IEEEauthorrefmark{3}ICREA Acadèmia, Spain\\
% \{mmartinez, aperez\}@cttc.es
% }}

\name{Marc Martinez-Gost $^{\star}$ 
\qquad Miguel Ángel Lagunas $^{\dagger}$
\qquad Ana Pérez-Neira $^{\star \dagger}$ 
\thanks{This work is part of the project SOFIA PID2023-147305OB-C32 funded by MICIU/AEI/10.13039/501100011033 and FEDER/UE.}
}

\address{$^{\star}$ Centre Tecnològic de Telecomunicacions de Catalunya, Spain \\
$^{\dagger}$ Dept. of Signal Theory and Communications, Universitat Politècnica de Catalunya, Spain\\
}

\include{acronyms}

\maketitle
\begin{abstract}
Type-based multiple access (TBMA) is a digital over-the-air computation (OAC) scheme that exploits symbol collisions over the wireless multiple-access channel to construct a histogram of the transmitted data at the receiver. This paper addresses the optimal detection of the transmitted histogram in TBMA over an additive white Gaussian noise (AWGN) channel. Since the number of possible received histograms grows combinatorially with both the number of transmitters and the data alphabet size, maximum-likelihood (ML) detection renders computationally infeasible. We prove that the ML detection problem can be solved by a greedy algorithm based on successive interference cancellation (SIC). We derive the analytical mean squared error (MSE) of the ML detector and show that it decreases exponentially with the SNR. Numerical results validate the theoretical analysis and demonstrate that TBMA achieves superior performance compared with conventional analog OAC, particularly for nonlinear functions.
\end{abstract}

\begin{keywords}
Over-the-air computation, Type-based multiple access, digital modulations.
\end{keywords}

\section{Introduction}

Over-the-air computation (OAC) \cite{sp_magazine, ota_original} exploits the superposition property of the wireless multiple-access channel to compute functions of distributed data directly over the air, avoiding the need to separately decode the data transmitted by each device. This paradigm is particularly attractive for networks in which a server must efficiently aggregate distributed measurements (e.g., wireless sensor networks) or locally computed values (e.g., federated learning).

Among the different approaches to OAC, type-based multiple access (TBMA) \cite{tbma_tong} has emerged as a prominent candidate. By exploiting the multiple-access channel, TBMA constructs a histogram of the transmitted data directly at the receiver, enabling the computation of a broad class of functions. Although TBMA requires greater transmission resources than foundational OAC schemes, it offers several practical benefits: it can leverage conventional digital modulation formats, such as M-ary \gls{FSK}, and maintain constant transmit power \cite{tbma_fsk, sahin_fsk}; nonlinear functions can be accommodated without modifying the underlying transmission scheme, in contrast to conventional OAC approaches based on nomographic decompositions, which might not exist for all functions; moreover, its aggregation mechanism provides natural robustness against adversarial strategies \cite{tbma_robust}.

A fundamental challenge, however, arises at the receiver during the detection stage. The number of possible received histograms grows combinatorially with the number of devices and data alphabet (i.e., histogram bins). Consequently, exhaustive-search maximum-likelihood (ML) detection quickly becomes computationally prohibitive and, for this reason, existing works often restrict TBMA to scenarios with a small number of devices or a limited data alphabet \cite{sahin2026generic}.

In this work, we show that ML detection of the received histogram over an AWGN channel can be solved by a greedy algorithm whose complexity scales linearly with both parameters. The proposed procedure can be interpreted as a simple extension of successive interference cancellation (SIC). We further derive the analytical mean squared error (MSE) achieved by the proposed detector and show that it exhibits an exponential decay with the signal-to-noise ratio (SNR). Finally, we compare the proposed approach with alternative OAC methods and demonstrate its superior performance.

\section{System Model}
Consider a wireless network consisting of $K$ distributed devices (transmitters) and a single server (receiver). Each device $k$ has a local data symbol $d_k$, assumed to be independent and identically distributed (i.i.d.), and satisfy \mbox{$-A< d_k < A$}.

Since TBMA is a digital communication scheme, the data must first be quantized into discrete values. Without loss of generality we consider an uniform quantizer, $\mathcal{Q}:d_k\rightarrow \bar{d}_k\in[1,\dots,N]$. Then, each transmitter maps $\bar{d}_k$ to an \mbox{$N$-dimensional} communication symbol \mbox{$\mathbf{s}_k=[s_{k,1},\dots,s_{k,N}]^T$} as follows,
\begin{equation}
    s_{k,n}=
    \begin{cases}
        1, & \bar{d}_k=n,\\
        0, & \bar{d}_k\neq n.
    \end{cases}
    \label{eq:tbma_comm_symbol}
\end{equation}
In this way, TBMA adopts an $N$-ary orthogonal signaling scheme, where each quantized value is associated with a distinct communication dimension (e.g., frequency tone).

The communication symbol is then mapped to a continuous-time baseband waveform $x_k(t)$, over a symbol interval $T$,
\begin{align}
    x_k(t)=\sum_{n=1}^Ns_{k,n}\varphi_n^{}(t)= \varphi_{\bar{d}_k}^{}(t)
    \quad
    0\leq t<T,
    \label{eq:tx_signal}
\end{align}
where $\{\varphi_1^{}(t),\dots,\varphi_N^{}(t)\}$ are orthonormal basis functions.

For OAC, we assume that all devices transmit simultaneously with ideal time and phase synchronization. This assumption is commonly adopted in the theoretical analysis of OAC, as it allows us to isolate the effects of the wireless channel from those of synchronization impairments. In practice, such synchronization can be achieved using appropriate synchronization techniques \cite{sahin_sync}. Furthermore, we assume that each device performs channel inversion \cite{power_control}. This requires channel state information (CSI) at the transmitters and is a common approach in OAC to achieve coherent superposition. Since the receiver observes a superposition of the signals, it cannot compensate for the individual channel coefficients after reception.

Then, the received signal is
\begin{align}
    y(t)&=\sum_{k=1}^K \sqrt{P_T^{}}x_k(t)+w(t)\nonumber\\
    &=\sqrt{P_T}\sum_{n=1}^N K_n\varphi_n^{}(t)+w(t)\quad
    0\leq t<T,
    \label{eq:rx_signal_inversion}
\end{align}
where $P_T^{}$ denotes the transmit power, $w(t)$ is complex baseband additive white Gaussian noise (AWGN) with spectral density $N_0/2$ and $K_n$ denotes the number of devices transmitting over resource $n$. Thus, \gls{TBMA} enables over-the-air superposition when users have identical discretized symbols.

The signal after matched filtering with $\varphi_n^{}(t)$ at sampling time and symbol time $t=T$ is
\begin{equation}
    r_n=\sqrt{P_T}K_n +\nu_n,
    \label{eq: MF_tbma}
\end{equation}
where $\nu_n$ is AWGN with spectral density $N_0/2$. This signal is a sufficient statistic of the number of transmitting devices per resource,
\begin{equation}
    \tilde{K}_n=\frac{r_n}{\sqrt{P_T}}\sim\mathcal{N}\left(
    K_n,\frac{N_0}{2P_T}
    \right),
    \label{eq:users_bin_hat}
\end{equation}

The received vector $(\tilde{K}_1,\dots,\tilde{K}_{N})$ can be interpreted as a noisy histogram of the quantized measurements $\bar{d}_k$.

The objective of the network is to compute a function of the distributed data, $f(d_1,\dots,d_K)$. In TBMA an estimate $\hat{f}$ is computed from the received histogram. For the theoretical derivations, we focus on the sample mean function,
\begin{equation}
    f(d_1,\dots,d_K)=\frac{1}{K}\sum_{k=1}^K d_k,
    \label{eq:mean_function}
\end{equation}
and the computation of nonlinear functions is considered subsequently in the simulation section. %Note that, although the data symbols $d_k$ may be modeled according to an underlying probability distribution, the function in \eqref{eq:mean_function} represents the sample mean of the $K$ realized data symbols and not the statistical mean of their distribution. For instance, with $K=1$ user, the function returns $d_1$. \mmg{remove this?}

The performance metric considered throughout this work is the mean squared error (MSE),
%%%%%%%%%%%%%%%%%%%%%%%%%%%
\begin{equation}
\text{MSE} = \mathbb{E}\bigg\{\left(f(d_1,\dots,d_K)-\hat{f}(d_1,\dots,d_K)\right)^2\bigg\},
\label{eq:figure_merit_mse}
\end{equation}
%%%%%%%%%%%%%%%%%%%%%%%%%%%
where the expectation is taken over the channel noise.

%Next, we describe the implementation of TBMA, including the modulation and demodulation procedures and the corresponding function-estimation stage at the receiver. \mmg{remove or rewrite}

\section{Maximum Likelihood Detection}
\label{sec:tbma}

From a detection-theoretic perspective, recovering the transmitted histogram corresponds to jointly detecting the vector $(\hat{K}_1,\ldots,\hat{K}_N)$ from the noisy observation $(\tilde{K}_1,\ldots,\tilde{K}_N)$. According to \eqref{eq:users_bin_hat}, the observation is corrupted by AWGN with independent components. Therefore, the ML detector reduces to the minimum Euclidean distance rule:
%--------------
\begin{subequations}
\begin{align}
\nonumber
& \underset{k_1,\dots,k_N}{\text{minimize}} ~~~~~~
\,\sum_{n=1}^N J_n(k_n)=\sum_{n=1}^N (\tilde{K}_n-k_n)^2 \\
& \hspace{-20pt}\qquad\text{subject to}~~~~~~
\sum_{n=1}^N k_n=K\label{const:sum_k_1}\\
&~~~~~~~~~~~~~~~~~~
k_n\in\mathbb{Z}_+,\forall n\label{const:nonneg}
\end{align}
\label{eq:min_mse}
\end{subequations}
%--------------
where constraint \eqref{const:sum_k_1} enforces the allocation of exactly $K$ users across $N$ resources, and constraint \eqref{const:nonneg} enforces nonnegative integer occupancies.

Since the number of possible received histograms increases exponentially with $K$ and $N$, the exhaustive search required by ML detection is computationally prohibitive. %That said, in the sequel we show that the ML detector admits a greedy implementation whose complexity scales linearly in each $N$ and $K$.

\subsection{Low-complexity algorithm}
We prove that the ML detector admits a greedy implementation whose complexity scales linearly in each $N$ and $K$.

Let $G_{n,k}$ denote the reduction in the objective function obtained when increasing the occupancy of resource $n$ from $k-1$ to $k$ users,
\begin{equation}
    G_{n,k}=J_n(k_n-1)-J_n(k_n)=2(\tilde{K}_n-k)+1,
    \label{eq:gain}
\end{equation}
which decreases linearly with the number of users. Equation \eqref{eq:gain} can be recursively expressed as
\begin{align}
    J_n(k_n)&=J_n(k_n-1) - G_{n,k}
    = J_n(0) - \sum_{i=1}^{k_n} G_{n,i},
    \label{eq:gain_recursive}
\end{align}
that is, the initial cost corresponding to zero allocated users minus the accumulated marginal gains.

Next, we introduce binary variables $z_{n,k}\in\{0,1\}$, where $z_{n,k}=1$ indicates that resource $n$ is assigned $k$ users. Then, the cost in \eqref{eq:min_mse} can be written as
\begin{equation}
    \sum_{n=1}^N J_n(k_n)=\sum_{n=1}^N J_n(0) -
    \sum_{n=1}^N \sum_{k=1}^K z_{n,k}G_{n,k}
    \label{eq:cost_gains}
\end{equation}

Since the first term in \eqref{eq:cost_gains} is constant, minimizing the Euclidean distance is equivalent to maximizing the accumulated gain. Therefore, problem \eqref{eq:min_mse} can be reformulated as
%--------------
\begin{subequations}
\begin{align}
\nonumber
& \underset{\{z_{n,k}\}}{\text{maximize}} ~~~~~~
\,\sum_{n=1}^N\sum_{k=1}^K z_{n,k}G_{n,k} \\
& \hspace{-20pt}\qquad\text{subject to}~~~~~~
\sum_{n=1}^N\sum_{k=1}^K z_{n,k}=K\label{const:sum_k_2}\\
&~~~~~~~~~~~~~~~~~~
z_{n,k}\in\{0,1\},\forall n,k\label{const:bool}\\
&~~~~~~~~~~~~~~~~~~
z_{n,k+1}\leq z_{n,k},\forall n,k\label{const:decrease}
\end{align}
\label{eq:max_gain}
\end{subequations}
%-------------- 
\vspace{-10pt}

Constraints \eqref{const:sum_k_2} and \eqref{const:bool} replace \eqref{const:sum_k_1} and \eqref{const:nonneg}, respectively. Constraint \eqref{const:decrease} enforces the natural ordering of the gains: selecting the $k$th gain of a given bin implies that all preceding gains from the same bin have already been selected. In other words, assigning the $k$th user to a bin necessarily implies the presence of the preceding $k-1$ users. This is implicit in the solution because the gains satisfy
\begin{equation}
    G_{n,1}>G_{n,2}>\dots>G_{n,K}
    \label{eq:monotone_gains}
\end{equation}

Although \eqref{eq:max_gain} is a combinatorial problem, the monotonicity of \eqref{eq:monotone_gains} makes it exactly solvable by a greedy procedure. For exposition, collect all possible gains into a matrix,
\begin{equation}
    \mathbf{G}=
     2\begin{bmatrix}
        \tilde{K}_1-1 & \tilde{K}_1-2 & \tilde{K}_1-3 & \cdots \\
        \tilde{K}_2-1 & \tilde{K}_2-2 & \tilde{K}_2-3 & \cdots \\
        \vdots & \vdots & \vdots & \\
        \tilde{K}_N-1 & \tilde{K}_N-2 & \tilde{K}_N-3 & \cdots \\
    \end{bmatrix}+1
    \label{eq:gain_matrix}
\end{equation}

Initially, only the first column of $\mathbf{G}$ is eligible for selection, since it contains the largest gain for each resource $n$. The largest entry corresponds to
\begin{equation}
    n^\star= \argmax_n \tilde{K}_n,
\end{equation}
which assigns one user to resource $n^\star$. Because of \eqref{eq:monotone_gains}, the selected entry is replaced by the succeeding gain in the same row, corresponding to assigning an additional user to that bin, while all other rows remain unchanged. This procedure is repeated until $K$ gains have been selected. Since every row is strictly decreasing, selecting a later gain from a row is never beneficial before selecting all preceding gains from the same row. Consequently, the greedy procedure always selects the globally largest feasible gain at each iteration and therefore solves \eqref{eq:max_gain} exactly.

%From an implementation viewpoint, the gains need not be computed explicitly. According to \eqref{eq:gain}, maximizing the gain $G_{n,k}$ is equivalent to maximizing the residual occupancy $\tilde{K}_n-k_n$. Consequently, each iteration simply identifies the largest histogram bin, assigns one user to the corresponding resource, and subtracts one unit from that histogram bin. Figure \ref{fig:SIC_decoder} illustrates this procedure with an example.

% \begin{figure*}[t]
%     \centering
%     \includegraphics[width=\textwidth]{img/histogram_SIC.pdf}
%   \caption{Example of the proposed SIC decoder over three iterations. At each iteration, the decoder identifies the histogram bin with the largest value, $n^\star$, decodes one user assigned to that bin, and subtracts one unit from the corresponding histogram entry. After three iterations, the recovered resource indices are $\bar{s}_k=\{3,3,5\}$.}
%   \label{fig:SIC_decoder}
% \end{figure*}

The greedy algorithm also admits a natural signal processing interpretation: At each iteration, the algorithm identifies the orthogonal resource with the largest signal power, assigns one user to that resource, and subtracts its contribution from the received histogram. The procedure is equivalent to successive interference cancellation (SIC), in which one signal component is detected and removed before proceeding to the next iteration \cite[Chapter 7]{verdu1998multiuser}. Equivalently, the algorithm can be interpreted as matching pursuit \cite{MP_comms}.

%The greedy selection further admits an intuitive communication perspective. According to \eqref{eq: MF_tbma}, the received signal power at each resource is proportional to the number of devices transmitting over that resource. Consequently, the histogram bin with the largest value corresponds to the resource with the highest effective SNR. Since higher SNR generally implies a lower detection error probability, it is natural to begin the decoding process from the largest histogram component. %This iterative detect-and-cancel strategy is precisely the principle underlying SIC.

%Unlike conventional SIC receivers employed in wireless communications, the proposed SIC is ML-optimal. The reason is that each detected component has a known and deterministic amplitude, so no amplitude estimation is required and each cancellation step exactly removes one unit contribution from the corresponding histogram bin. Therefore, the marginal gains remain ordered as in \eqref{eq:gain}, ensuring that the greedy procedure coincides with the ML solution. Furthermore, the receiver does not attempt to identify the device associated with each recovered contribution. This is unnecessary in OAC, since the objective is to estimate a permutation-invariant function of the transmitted measurements rather than to recover the individual transmitters. Thus, recovering the measurements in arbitrary order is sufficient for accurate function computation.

\subsection{Computation and performance}
Any symmetric function can be computed from the clean histogram $(\hat{K}_1,\dots, \hat{K}_N)$. For instance, the sample mean is computed as
\begin{equation}
    \hat{f}=\frac{1}{K}\sum_{n=1}^N \mathcal{Q}^{-1}(n)\hat{K}_n,
\end{equation}
which corresponds to the first moment, and $\mathcal{Q}^{-1}(n)$ maps the histograms bins back to the original data domain.

To evaluate the theoretical performance on the sample mean function, we acknowledge that for sufficiently large $N$, the quantization distortion becomes uncorrelated with the data. It follows that
\begin{align}
    \text{MSE}&=\underbrace{\mathbb{E}\Big\{\left(f-\bar{f}\,\right)^2\Big\}}_{\text{MSE}_\text{q}}+
    \underbrace{\mathbb{E}\Big\{\left(\bar{f}-\hat{f}\right)^2\Big\}}_{\text{MSE}_\text{c}},
    \label{eq:independence_mse}
\end{align}
where $\bar{f}=\sum_k\bar{d}_k/K$, $\text{MSE}_\text{q}$ correspond to the quantization distortion and $\text{MSE}_\text{c}$ to channel-induced distortion.

The quantization error can be modeled as an independent uniformly distributed random variable over one quantization interval, which results in
\begin{align}
    \text{MSE}_\text{q}= \mathbb{E}\Bigg\{\left(\frac{1}{K}\sum_{k=1}^K d_k-\frac{1}{K}\sum_{k=1}^K\hat{d}_k\right)^2\Bigg\}
    =\frac{A^2}{3N^2K}
    \label{eq:mse_q}
\end{align}

Since the quantization distortion is unaffected by the decoding strategy, we focus exclusively on the channel-induced component. This contribution corresponds to
\begin{align}
    \text{MSE}_\text{c}&=
    \mathbb{E}\bigg\{
    \left(\bar{f}-\hat{f}\right)^2\Big|
    \hat{K}_n\neq K_n,\forall n
    \bigg\}
    \text{Pr}
    \left(
    \hat{K}_n\neq K_n,\forall n
    \right),
    \label{eq:mse_conditioned}
\end{align}
and the error is zero when the histogram is decoded correctly.

To derive an analytical expression, we consider the high-SNR regime. As in conventional digital communications, decoding errors are then dominated by nearest-neighbor events. Thus, the histogram error probability can be upper bounded by the union bound and the pairwise error probability,
\begin{equation}
    \text{Pr}
    \left(
    \hat{K}_n\neq K_n,\forall n
    \right)\leq
    N_\text{nn}Q
    \left(
    \sqrt{\frac{d_\text{min}^2}{4\sigma_w^2}}
    \right),
    \label{eq:SER_histogram}
\end{equation}
where $N_\text{nn}$ denotes the number of nearest-neighbor histograms, $d_\text{min}$ is their Euclidean distance and $\sigma_w^2$ is the noise variance of $\tilde{K}_n$ in \eqref{eq:users_bin_hat}.

When symbols are histograms, a nearest-neighbor corresponds to moving a single user from one occupied bin to another. Without loss of generality, consider a user moved from the first to the second bin, yielding the minimum Euclidean distance
\begin{equation}
    d_\text{min}=||(-1,1,0,\dots,0)||=\sqrt{2}
    \label{d_min}
\end{equation}

The number of nearest neighbors is obtained by selecting one occupied bin as the original user bin and one of the remaining $N-1$ bins as the destination. We upper-bound the number of occupied bins with $\min(K,N)$: When $K<N$, the maximum is attained when every user occupies a different bin, whereas for $K\geq N$ all bins can be occupied. Thus, we can upper-bound the number of nearest neighbors by
\begin{equation}
    N_\text{nn}\leq (N-1)\min(K,N)
    \label{eq:neighbors}
\end{equation}
Substituting \eqref{d_min} and \eqref{eq:neighbors} into \eqref{eq:SER_histogram} gives
\begin{equation}
    \text{Pr}
    \left(
    \hat{K}_n\neq K_n,\forall n
    \right)\leq
    N\min(K,N)Q
    \left(
    \sqrt{\frac{E_T}{N_0}}
    \right),
    \label{eq:SER_histogram2}
\end{equation}
with $E_T^{}=TP_T^{}$ being the total energy per transmitter.

The remaining component is the conditional error in \eqref{eq:mse_conditioned}. Under the assumption of a nearest-neighbor error, the MSE corresponds to
\begin{align}
    % \mathbb{E}\bigg\{
    % \left(\bar{f}-\hat{f}\right)^2\Big|
    % \hat{K}_n\neq K_n,\forall n
    % \bigg\}&=\\
    \mathbb{E}\bigg\{
    \left(\bar{f}-\hat{f}\right)^2\Big|
    \bar{d}_1\neq\mathcal{Q}\left(\hat{d}_1\right)
    \bigg\}\nonumber
    &=\frac{A^2}{K^2}\mathbb{E}\left\{\left(
    n-\bar{d}_1
    \right)^2\right\}\nonumber\\
    &=\frac{2A^2}{3K^2},
    \label{eq:mse_bound2}
\end{align}
where the expectation is taken over all $N$ bins. Specifically, under orthogonal signaling, an erroneous symbol is equally likely to be decoded as any other symbol. Thus, the last equality follows from the variance of a uniform distribution. 

It is worth emphasizing that orthogonal transmissions across users does not provide the same error behavior. If the users were instead transmitted one at a time, the receiver would perform $K$ separate detections and the resulting function estimate would contain $K$ independent noise contributions. In contrast, with TBMA, the channel-induced MSE is governed by a single error event.

Finally, substituting \eqref{eq:SER_histogram2} and \eqref{eq:mse_bound2} into \eqref{eq:mse_conditioned} yields the following the channel-induced MSE at high-SNR:
\begin{equation}
    \text{MSE}_\text{c}\approx
    \frac{2A^2N\min(K,N)}{3K^2}
    Q
    \left(
    \sqrt{\frac{E_T}{N_0}}
    \right)
\end{equation}

\section{Numerical Results}
In this section, we evaluate the performance of the proposed ML detector and compare it with state-of-the-art methods. We consider $d_k\sim\text{Uniform}(-1,1)$ and use MFSK as modulation.

\begin{figure}[t]
    \centering
    \includegraphics[width=\columnwidth]{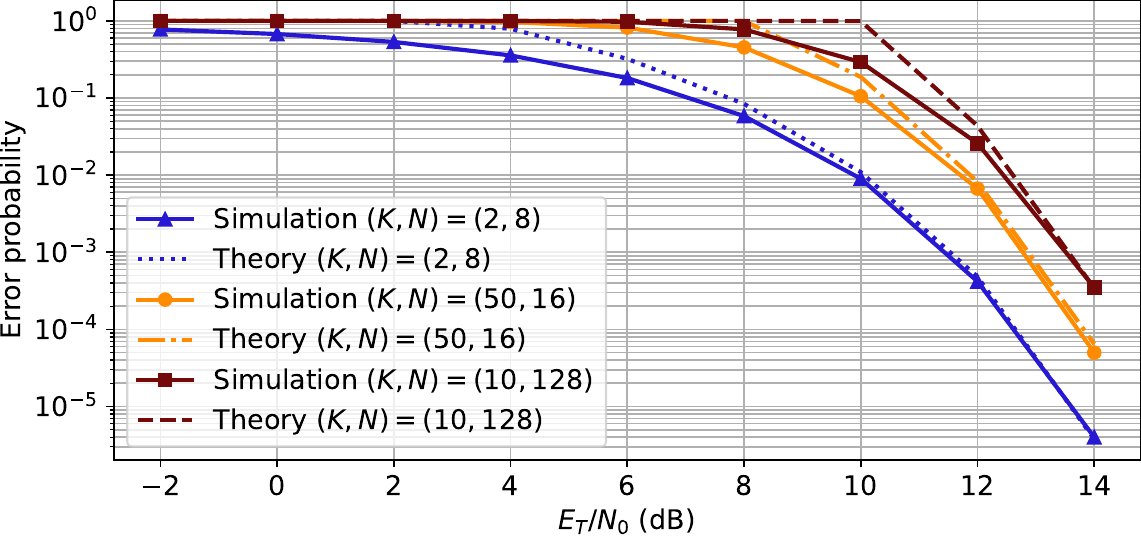}
    \caption{Theoretical and simulated error probabilities for histogram detection under different $(K,N)$ configurations.}
    \label{fig:error_prob}
\end{figure}

Figure \ref{fig:error_prob} illustrates the error probability in histogram detection obtained through simulations with the proposed SIC detector, along with the theoretical bound, for different values of $K$ and $N$. The results show that the bound provides a good approximation for $E_T/N_0 \geq 12$ dB, which can be considered the high-SNR regime.

\begin{figure}[t]
    \centering
    \includegraphics[width=\columnwidth]{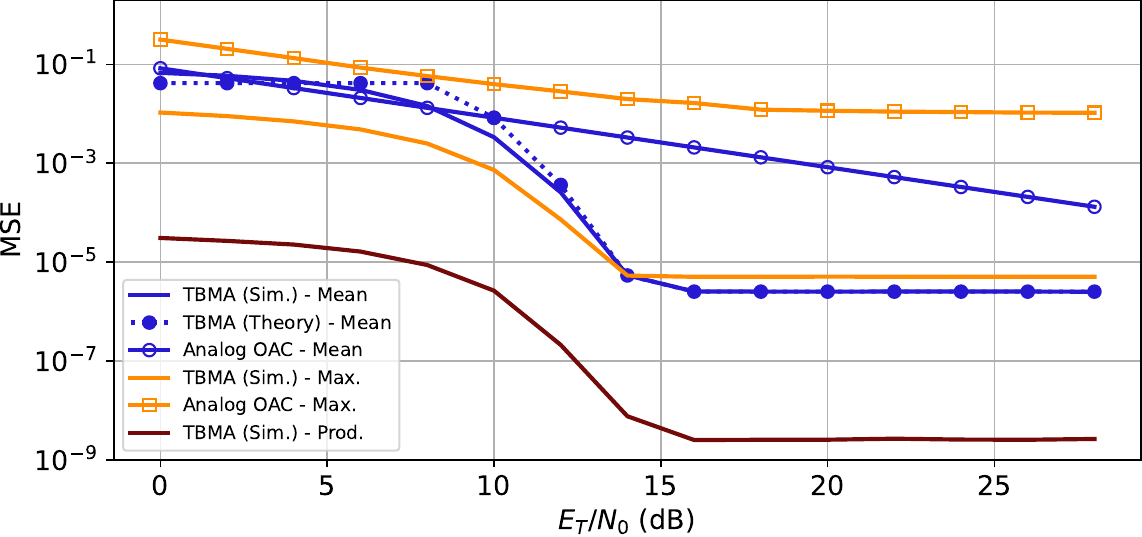}
    \caption{Theoretical and simulated MSE for TBMA and analog OAC across different functions, with $(K,N)=(10,128)$.}
    \label{fig:mse}
\end{figure}

Figure \ref{fig:mse} illustrates the MSE for different functions. We compare TBMA with conventional analog OAC with amplitude modulation (see \cite{sp_magazine}). For the mean function, the MSE of analog OAC decreases linearly with the SNR, whereas that of TBMA decays exponentially until reaching the quantization floor. For the maximum function, TBMA exhibits a similar performance, while analog OAC degrades due to the approximation involved in its nomographic decomposition. For the product function, TBMA exhibits the same exponential decay, whereas analog OAC cannot be directly applied because no nomographic decomposition exists for products involving both positive and negative values. Overall, these results highlight the effectiveness of type-based OAC, specifically for nonlinear functions.

\section{Conclusion}
This paper presented a low-complexity ML detector for TBMA over an AWGN channel. While exhaustive search is infeasible, the optimal histogram can be detected using a greedy SIC-based procedure with complexity that scales linearly with the number of devices and the data alphabet size. We also derived the corresponding analytical MSE for the mean function and showed that it decreases exponentially with the SNR. Comparisons with conventional analog OAC demonstrated the advantages of TBMA, particularly for nonlinear functions. Overall, these results highlight TBMA as a flexible and general framework for OAC.

\bibliographystyle{IEEEbib}
\bibliography{refs}

\end{document}

%% file: acronyms.tex
\newacronym{AI}{AI}{Artificial Intelligence}
\newacronym{AirComp}{OAC}{over-the-air computation}
\newacronym{AWGN}{AWGN}{additive white Gaussian noise}
\newacronym{CNN}{CNN}{convolutional neural network}
\newacronym{CSI}{CSI}{channel state information}
\newacronym{DA}{DA}{direct aggregation}
\newacronym{DSB}{DSB}{double sideband}
\newacronym{FL}{FEEL}{federated learning}
\newacronym{FSK}{FSK}{frequency shift keying}
\newacronym{MSE}{MSE}{mean squared error}
\newacronym{NMSE}{NMSE}{normalized mean squared error}
\newacronym{PAM}{PAM}{pulse amplitude modulation}
\newacronym{PPM}{PPM}{pulse position modulation}
\newacronym{TBMA}{TBMA}{type-based multiple access}
\newacronym{SNR}{SNR}{signal-to-noise ratio}

%% file: refs.bib
@ARTICLE{ota_original,
  author={Nazer, Bobak and Gastpar, Michael},
  journal={IEEE Transactions on Information Theory}, 
  title={Computation Over Multiple-Access Channels}, 
  year={2007},
  volume={53},
  number={10},
  pages={3498-3516},
  doi={10.1109/TIT.2007.904785}}

@ARTICLE{sp_magazine,
  author={Pérez-Neira, Ana and Martinez-Gost, Marc and Şahin, Alphan and Razavikia, Saeed and Fischione, Carlo and Huang, Kaibin},
  journal={IEEE Signal Processing Magazine}, 
  title={Waveforms for Computing Over the Air: A groundbreaking approach that redefines data aggregation}, 
  year={2025},
  volume={42},
  number={2},
  pages={57-77},
  doi={10.1109/MSP.2024.3500775}}

@ARTICLE{tbma_tong,
  author={Mergen, G. and Tong, L.},
  journal={IEEE Transactions on Signal Processing}, 
  title={Type based estimation over multiaccess channels}, 
  year={2006},
  volume={54},
  number={2},
  pages={613-626},
  doi={10.1109/TSP.2005.861896}}

@inproceedings{sahin_fsk,
  title={Distributed learning over a wireless network with FSK-based majority vote},
  author={{\c{S}}ahin, Alphan and Everette, Bryson and Hoque, Safi Shams Muhtasimul},
  booktitle={2021 4th International Conference on Advanced Communication Technologies and Networking (CommNet)},
  pages={1--9},
  year={2021},
  organization={IEEE}
}

@INPROCEEDINGS{tbma_fsk,
  author={Martinez-Gost, Marc and Pérez-Neira, Ana and Lagunas, Miguel Ángel},
  booktitle={GLOBECOM 2023 - 2023 IEEE Global Communications Conference}, 
  title={Frequency Modulation Aggregation for Federated Learning}, 
  year={2023},
  volume={},
  number={},
  pages={1878-1883},
  doi={10.1109/GLOBECOM54140.2023.10437413}}

@INPROCEEDINGS{tbma_robust,
  author={Martinez-Gost, Marc and Pérez-Neira, Ana and Lagunas, Miguel Ángel},
  booktitle={2025 33rd European Signal Processing Conference (EUSIPCO)}, 
  title={Robust Over-the-Air Computation with Type-Based Multiple Access}, 
  year={2025},
  volume={},
  number={},
  pages={2042-2046},
  doi={10.23919/EUSIPCO63237.2025.11226299}}

@article{sahin2026generic,
  title={A Generic Multi-dimensional Symbol Construction for Digital Over-the-Air Computation and Practical Aspects},
  author={Sahin, Alphan},
  journal={arXiv preprint arXiv:2606.18085},
  year={2026}
}

@inproceedings{sahin_sync,
  title={On the feasibility of distributed phase synchronization for coherent signal superposition},
  author={{\c{S}}ahin, Alphan},
  booktitle={2025 IEEE 36th International Symposium on Personal, Indoor and Mobile Radio Communications (PIMRC)},
  pages={1--6},
  year={2025},
  organization={IEEE}
}

@ARTICLE{power_control,
  author={Cao, Xiaowen and Zhu, Guangxu and Xu, Jie and Huang, Kaibin},
  journal={IEEE Transactions on Wireless Communications}, 
  title={Optimized Power Control for Over-the-Air Computation in Fading Channels}, 
  year={2020},
  volume={19},
  number={11},
  pages={7498-7513},
  doi={10.1109/TWC.2020.3012287}}

@book{verdu1998multiuser,
  title={Multiuser detection},
  author={Verdu, Sergio},
  year={1998},
  publisher={Cambridge university press}
}

@ARTICLE{MP_comms,
  author={Cotter, S.F. and Rao, B.D.},
  journal={IEEE Transactions on Communications}, 
  title={Sparse channel estimation via matching pursuit with application to equalization}, 
  year={2002},
  volume={50},
  number={3},
  pages={374-377},
  doi={10.1109/26.990897}}
